\documentclass[twocolumn,showpacs,pra,aps,superscriptaddress]{revtex4}

\usepackage{graphicx}

\begin{document}

\title{Full time nonexponential decay in double-barrier quantum structures}

\author{Gast\'on Garc\'{\i}a-Calder\'on}
\email{gaston@fisica.unam.mx} \affiliation{Departamento de Qu\'{\i}mica-F\'{\i}sica, Universidad del Pa\'{\i}s Vasco Apartado Postal 644,
Bilbao, Spain} \altaffiliation{Permanent address: Instituto de F\'{\i}sica, Universidad Nacional Aut\'onoma de M\'exico, Apartado Postal {20
364}, 01000 M\'exico, D.F., M\'exico}

\author{Jorge Villavicencio}
\email{villavics@uabc.mx} \affiliation{Facultad de Ciencias, Universidad Aut\'onoma de Baja California, Apartado Postal 1880, 22800 Ensenada,
Baja California, M\'exico}

\date{\today}

\begin{abstract}
We examine an analytical expression for the survival probability for the time evolution of quantum decay to discuss a regime where quantum decay
is nonexponential at all times. We find that the interference between the exponential and nonexponential terms of the survival amplitude
modifies the usual exponential decay regime in systems where the ratio of the resonance energy to the decay width, is less than $0.3$. We
suggest that such regime could be observed in semiconductor double-barrier resonant quantum structures with appropriate parameters.
\end{abstract}


\maketitle

\section{Introduction}

The exponential decay law has been very successful in describing the time evolution of quantum decay \cite{gamow,gurney}. However, almost 50
years ago, Khalfin \cite{khalfin} showed that for quantum systems whose energy spectra is bounded from below, \textit{i.e.}, $(0,\infty)$, which
encompasses the vast majority of systems found in Nature, the exponential decay law cannot hold in the full time interval. The present commonly
accepted view of the time evolution of decay involves three clearly distinguishable time regimes in terms of the lifetime of the system
\cite{exner,onley,muga,greenland,dicus,rufeil}: (a) Nonexponential decay at short times, (b) Exponential behavior spanning over many lifetimes
at intermediate times, (c) Nonexponential decay as an inverse power law of time at long times. The experimental confirmation of the
nonexponential behavior has remained elusive over decades. After years of experimental effort, dealing mainly with radioactive atomic nucleus
\cite{norman}, and elementary particles \cite{opal}, the deviation from the exponential decay law in the short-time limit has been finally
reported some years ago for an artificial quantum system \cite{raizen}. In the framework of an exact single resonance decay model \cite{gcrr01},
it is illustrated that the deviation at long times depends on the value of the ratio of the resonance energy $\varepsilon_r$ to the decay width
$\Gamma_r$, \textit{i.e.}, $ R=\varepsilon_r/\Gamma_r$ \cite{threshold}. As the value of $R$ diminishes, from very large values up to values of
the order of unity,  the long time deviation from exponential decay occurs earlier as a function of the lifetime, $\tau=\hbar/\Gamma_r$, of the
corresponding system. In a recent work, Jittoh \textit{et al.} \cite{jittoh}, have shown that for values of $R$ still smaller \textit{i.e.},
less unity, there exists a \textit{novel} regime where the decay is nonexponential at all times. These authors left the discussion of full time
nonexponential decay in actual physical systems for future work.

In this work we consider an analytical expression for the time evolution of decay for finite range potentials to discuss further the regime of
nonexponential decay in the full time interval.  We show that the absence of the exponential period in decay is due to the interference between
the exponential  and nonexponential contributions to decay. It is also suggested that one-dimensional semiconductor double-barrier resonant
quantum structures may be suitable systems to verify experimentally that behavior.

Section II presents the formalism, Section III deals with the calculations and its discussion, and Section IV gives the concluding remarks of
this work.

\section{Formalism}

Let us therefore consider the decay of an arbitrary state $\psi(x,t=0)$, initially confined at $t=0$, along the internal region $0 \le x \le L$
of a one-dimensional potential $V(x)$ that vanishes beyond a distance, \textit{i.e.} $V(x)=0$ for $ x \leq 0$ and $x \geq L$. The solution
$\psi(x,t)$ at time $t > 0$ may be expressed in terms of the retarded Green´s function $g(x,x';t)$ of the problem as,
\begin{equation}
\psi \left( x,t\right) =\int_{0}^{L} g(x,x^{\prime };t)\psi (x^{\prime},0)\,dx^{\prime }. \label{1}
\end{equation}
The survival  amplitude $A(t)$, that provides the probability amplitude that the evolved function $\psi(x,t)$ at time $t$ remains in the initial
state $\psi(x,0)$ is defined as,
\begin{equation}
A(t) =\int_{0}^{L}\psi ^{\ast}( x,0) \psi (x,t)\,dx, \label{2}
\end{equation}
and consequently the survival probability reads $S(t)=|A(t)|^2$. A convenient approach to solve Eq.\ (\ref{2}) is by Laplace Transforming
$g(x,x';t)$ into the momentum k-space to exploit the analytical properties of the outgoing Green function $G^{+}(x,x^{\prime};k)$ of the problem
\cite{newton}. Here we follow and generalize to one dimension the approach developed by Garc\'{\i}a-Calder\'on in three dimensions \cite{gc92}.

The essential point  of our approach is that the full outgoing Green's function $G^{+}(x,x^{\prime};k)$ of the problem  may be written as an
expansion involving its complex poles $\{k_n\}$ and residues, the resonance functions $\{u_n(x)\}$ \cite{gcr97}. We restrict the discussion to
potentials that do not hold bound nor antibound states. In general, the complex poles $k_n=\alpha_n-i\beta_n$ with $(\alpha_n,\beta_n)>0$,
$(n=1,2,...,)$,  are simple and are distributed along the lower-half of the $k$-plane in a well known manner\cite{newton}. From time-reversal
considerations, those seated on the third quadrant, $k_{-n}$, are related to those on the fourth, $k_n$, by $k_{-n}=-k_n^*$. Analogously, the
residues fulfil $u_{-n}=u_n^*$. The complex energy poles $E_n=\varepsilon_n-i\Gamma_n/2$ may be written in terms of $k_n$ as
$E_n=\hbar^2k_n^2/2m$ and hence  $\varepsilon_n=\hbar^2(\alpha_n^2-\beta_n^2)/2m$ and $\Gamma_n=\hbar^2(4\alpha_n\beta_n)/2m$, with $m$ the mass
of the particle. Since the energy of the decaying particle, $\varepsilon_n$, is necessarily positive, the poles of the system  must be the so
called \textit{proper resonance poles} \textit{i.e.}, poles satisfying $\alpha_n > \beta_n$. Note that this implies that $R > 0$. As a result of
the above considerations the survival amplitude may be expressed as a sum over exponential and nonexponential contributions, the latter being in
general relevant at very short and long times compared with the lifetime. Hence we write \cite{gc92},
\begin{eqnarray}
A(t)= \sum_{n=1}^{\infty}&&  \{ C_n \bar{C}_n e^{-i\hbar k_n^2t/2m}-\nonumber\\[.3cm]
&& [C_n \bar{C}_nM(-y_n) - (C_n \bar{C}_n)^*M(y_{-n})] \}, \label{3a}
\end{eqnarray}
where the function $M(y_q)$ is defined as,
\begin{equation}
M(y_q) =\frac{i}{2\pi }\int_{-\infty }^{\infty} \frac{e^{-i\hbar k^2t/2m}}{k-k_q}dk=\frac{1}{2}w( iy_q), \label{5}
\end{equation}
where $y_q=-\exp(-i \pi /4)(\hbar/2m)^{1/2}k_qt^{1/2}$, with $q= {\pm n}$, and the function $w(z)=\exp(-z^2)\rm{erfc(-iz)}$ is a well known
function \cite{abramowitz}. Proper resonance poles fulfil, $\pi/2 < \arg\, y_{n} < 3\pi/4$. The coefficients $C_{n}$ and $\bar{C}_n$ in
Eq.\,(\ref{3}) are given by,
\begin{equation}
C_{n}=\int_{0}^{L}\psi \left( x,0\right) u_{n}(x)\,dx;\, \bar{C}_{n}=\int_{0}^{L}\psi ^{\ast }\left( x,0\right)u_{n}(x)\,dx. \label{7}
\end{equation}
The above coefficients obey relationships that are similar to those in 3 dimensions \cite{gc92},
\begin{equation}
{\rm Re} \left (\sum_{n=1}^{\infty }C_{n}\bar{C}_{n}\right )=1,\,\,\, {\rm Im} \left( \sum_{n=1}^{\infty }\frac{C_{n}\bar{C}_{n}}{k_{n}}\right
)=0. \label{11}
\end{equation}
The resonant functions $u_n(x)$, necessary to calculate the coefficients given by Eq.\,(\ref{7}), satisfy the Schr\"odinger equation of the
problem  with complex  eigenvalues $k_n^2$. They obey outgoing boundary conditions at $x=0$ and $x=L$, given respectively by,
$[du_n(x)/dx]_{x=0}=-ik_n u_n(0)$, and $ [du_n(x)/dx]_{x=L}=i k_n u_n(L)$. Alternatively, the resonance functions can also be obtained from the
residues at the complex poles of $G^+(x,x';k)$ \cite{gcr97}. This yields a normalization condition that differs slightly from that in 3
dimensions, namely,
\begin{equation}
\int_0^Lu_n^2(x)dx +i {u_n^2(0)+u_n^2(L) \over 2k_n} =1. \label{3}
\end{equation}
The set of $\{k_n\}$'s and the corresponding $\{u_n\}$'s  that follow from the solution of the above complex eigenvalue problem, may be obtained
by well known methods\cite{gcr97}.

The long time behavior of Eq.\ (\ref{3a}), \textit{i.e.}, much larger than the lifetime $\tau=\hbar/\Gamma_1$,  rests only on the $M$-functions.
At long times they  behave as \cite{abramowitz} $M(y_q) \approx - a/(k_qt^{1/2}) - b/(k_q^3t^{3/2}) +..., $, with $q \pm n$, and the constants
$a = i/[2(\pi i)^{1/2}]$ and $b =1/[4(\pi i)^{1/2}]$. Substitution of the above expansion into Eq.\,(\ref{3a}) gives that the factor multiplying
$t^{-1/2}$ is proportional to the term given precisely by the expression on the right in Eq.\ (\ref{11}), and hence the $t^{1/2}$ contribution
vanishes exactly. This leads to the well known long time behavior of $A(t)$ as $t^{-3/2}$.

We shall be concerned here in situations where the initial state $\psi(x,0)$ overlaps strongly with the lowest energy resonant state $u_1(x)$ of
the system. In such a case it follows from the first expression  in Eq.\, (\ref{11}), that ${\rm Re}(C_1\bar {C}_1)$  is the dominant
contribution. Since the decaying widths and resonance energies satisfy, respectively, that $ 0 < \Gamma_1 < \Gamma_2 <...,$, and $\varepsilon_1
< \varepsilon_2 < ...,$, it follows that the higher resonance contributions decay much faster and may be neglected. This simplifies our
description of decay because it allows to deal with the single term approximation of the survival amplitude $A(t)$ given by Eq.\ (\ref{3a}).
This approximation also demands to make sure that the correct long-time behavior of the survival amplitude is preserved, namely $A(t) \sim
t^{-3/2}$. This requires to remove the $t^{-1/2}$  contribution in the $M's$ since it cancels out exactly in Eq.\ (\ref{3a}) \cite{gc92}. As a
consequence, the  single term approximation of the survival amplitude  may be written as,
\begin{equation}
A(t) =  A(t)_{exp} + A(t)_{non}, \label{12}
\end{equation}
where $A(t)_{exp}$ is,
\begin{equation}
A(t)_{exp}= C_1 \bar{C}_1 e^{-i\hbar k_1^2t/2m} \label{12a}
\end{equation}
and,
\begin{equation}
A(t)_{non}= C_1 \bar{C}_1{\cal M}(-y_1) - (C_1 \bar{C}_1)^*{\cal M}(y_{-1}), \label{12b}
\end{equation}
where the ${\cal M}'s$ denote $M$ functions where the long time contribution that goes as $t^{-1/2}$ has been subtracted to obtain the correct
long time behavior as $t^{-3/2}$. Hence the survival probability becomes
\begin{equation}
S(t)= S(t)_{exp}+S(t)_{non}+S(t)_{int}, \label{12c}
\end{equation}
where $S(t)_{exp}$, $S(t)_{non}$, and $S(t)_{int}$, refer respectively, to the exponential, nonexponential and interference contributions to the
survival probability, namely,
\begin{equation}
S(t)_{exp}= |C_1 \bar{C}_1|^2 e^{-\Gamma_1t/\hbar} \label{12d}
\end{equation}
\begin{equation}
S(t)_{non}=|A(t)_{non}|^2 \label{12e}
\end{equation}
and,
\begin{equation}
S(t)_{int}= 2 |C_1 \bar{C}_1|\cos (\varepsilon_1 t/\hbar + \eta+ \phi(t)) e^{-\Gamma_1t/2\hbar}|A(t)_{non}|, \label{12f}
\end{equation}
where  $\eta=\arg(C_1{\bar C}_1)$ and $\phi(t)=\arg(A(t)_{non})$. At long times, $A(t)_{non}$  may be written as an asymptotic expansion, that
we denote by $A(t)_{non}^{\ell}$, whose leading term reads \cite{gc92},
\begin{equation}
A(t)_{non}^{\ell} \approx -  \frac{e^{i\pi/4}}{2\sqrt{\pi}}\,\left(\frac{2m}{\hbar}\right)^{3/2}{\rm Im}\, \left \{\frac{C_1\bar{C}_1}{k_1^3}
\right \} \frac{1}{t^{3/2}}. \label{13}
\end{equation}
Consequently, at long times, $S(t)_{non}$ and $S(t)_{int}$, given respectively by Eqs.\ (\ref{12e}) and (\ref{12f}), may be written in obvious
notation as,
\begin{equation}
S(t)_{non}^{\ell} \approx  \frac{1}{4\pi}\,\left(\frac{2m}{\hbar}\right)^{3} {\rm Im}\,\left \{\frac{C_1\bar{C}_1}{k_1^3} \right \}^2
\frac{1}{t^{3}}. \label{16}
\end{equation}
and,
\begin{eqnarray}
S(t)_{int}^{\ell}& \approx &  -\frac{1}{\sqrt{\pi}}\,\left(\frac{2m}{\hbar}\right)^{3/2}|C_1\bar{C}_1|\,{\rm Im}\,
\left \{\frac{C_1\bar{C}_1}{k_1^3} \right \} \times \nonumber \\[.3cm]
&&\cos (\varepsilon_1 t/\hbar + \eta+\pi/4) e^{-\Gamma_1t/2\hbar}\frac{1}{t^{3/2}}. \label{17}
\end{eqnarray}

\section{Calculations and discussion}

In order to study systematically the behavior of the survival probability with time, in addition to the resonance parameters one needs to
specify the initial state. It is shown below, however, that if the one-term approximation holds, then the specific form of the initial state is
not essential to determine the behavior with time of the survival probability. Hence we choose for  $\psi(x,0)$ a very simple model, namely, the
box model state $\psi (x,0)=(2/w)^{1/2} \sin \pi(x-b)/w$ for $b  \le x \le b + w$ and zero, otherwise, where $b$ and $w$ stand respectively, for
the barrier and well widths. The corresponding box momentum $k=\pi/w$ is closer to the real part of the resonant momentum $\alpha_1$, than to
any other $\alpha$'s of the system.

\begin{figure}[!tbp]
\rotatebox{0}{\includegraphics[width=3.3in]{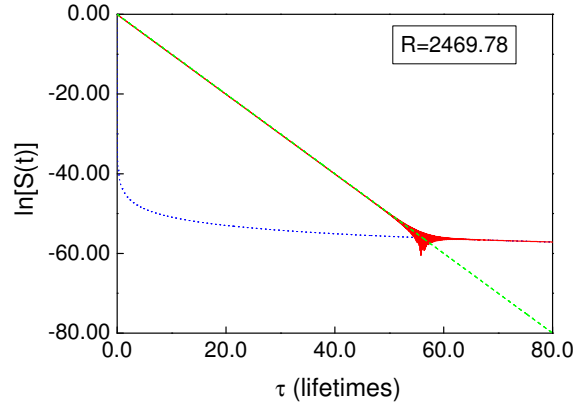}} \caption{ `online color'. The Survival probability $S(t)$ (solid line) as a function
of time for a double-barrier system with $R$ as indicated and $\tau_1=16321.9$ fs. The purely exponential behavior of $S(t)$ (dashed line) and
the long-time asymptotic behavior as $t^{-3}$ (dotted line), are included for comparison. See text.} \label{tsuchiya}
\end{figure}

In Fig.  \ref{tsuchiya}  we have used a set of potential parameters typical of  AlAs-GaAs-AlAs double-barrier heterostructures as in the cases
considered by Sakaki and co-workers \cite{sakaki}, who verified experimentally that electrons in sufficiently thin symmetric double-barrier
resonant structures decay proceeds according to the exponential decay law. The potential parameters are: barrier widths $b=2.5$ nm, well width
$w=6.2$ nm, and barrier heights $V=1.36$ eV. In the calculations the electron effective mass  is $m=0.067 m_e$, where $m_e$ is the bare electron
mass. The resonance parameters of the system are: resonance energy $\varepsilon_1=0.09959$ eV, resonance width $\Gamma_1=4.0325 \times 10^{-5}$
eV. Hence $R=\varepsilon_1/\Gamma_1=2469.78$, much larger than unity, and the lifetime is $\tau_1=\hbar/\Gamma_1=16321.9$ fs. The survival
probability $S(t)$ (solid line) is calculated using Eq.\ (\ref{12c}). For comparison, Fig. \ref{tsuchiya} exhibits also $S(t)^{\ell}_{non}$ (dot
line), given by Eq.\ (\ref{16}), and the purely exponential contribution $\exp(-\Gamma_1t/\hbar)$ (dashed line), \textit{i.e.}, Eq.\ (\ref{12d})
with $C_1\bar{C}_1=1$. We see that exponential decay law stands for many lifetimes. The long time nonexponential contribution  becomes  relevant
only after $60$ lifetimes when the value of $S(t)$  is extremely small, most possibly beyond experimental verification.

\begin{figure}[!tbp]
\rotatebox{0}{\includegraphics[width=3.3in]{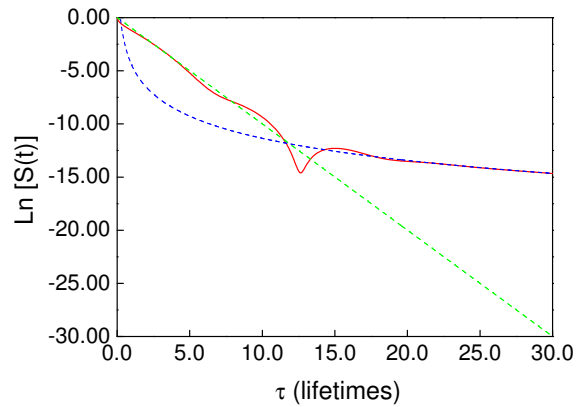}} \caption{`online color'. The Survival probability $S(t)$ (solid line), as a function
of time in lifetime units for a double-barrier system exhibiting the transition from exponential to nonexponential behavior. Here $\tau_1=10.69$
fs. The purely exponential behavior of $S(t)$ (dashed line) and the long-time asymptotic behavior (dotted line), are included for comparison.
See text.} \label{modsollner}
\end{figure}

The oscillations of the survival probability around the exponential-nonexponential transition in Fig. \ref{tsuchiya}, are caused by the cosine
factor appearing in the interference term $S_{int}$ given by Eq.\ (\ref{12f}). In the long time  limit, \textit{i.e.}, Eq.\ (\ref{17}), the
cosine factor may be expressed in terms of the parameters $R$ and $\tau$ as  $\cos(R\tau+\eta)$, and it explains the extremely large frequency
of oscillations of $S(t))$ around the exponential-nonexponential transition for $R >>1$.

By varying the potential parameters, one obtains also that the onset of the exponential-nonexponential transition in lifetime units, depends on
the value  $R=\varepsilon_1/\Gamma_1$. In fact, it occurs earlier as $R$ diminishes. This also occurs in the case of the exact single resonance
decay formula, whose only input is the value of $R$ \cite{gcrr01}. One sees, from the above cosine factor,  that the frequency of oscillations
diminishes also as $R$ becomes smaller. This is interesting, because it may allow to design structures with appropriate parameters to exhibit
nonexponential behavior in ranges more adequate for experiment.

As an example of this, Fig. \ref{modsollner} exhibits a plot of the survival probability $S(t)$ (solid line) in a case where $R=0.91$. The
potential  parameters of the double-barrier structure are \cite{ferry}, barrier widths $b=1.0$ nm, well width $w=5.0$ nm, and barrier heights
$V=0.23$ eV which give: $\varepsilon_1=0.05639$ eV, $\Gamma_1=0.06151$ eV and $\tau_1=10.69$ fs. The nonexponential behavior is set now around
$15$ lifetimes and the value of $S(t)$ is order of magnitudes larger than in the preceding case. Again, for comparison, $S(t)^{\ell}_{non}$ (dot
line) and the purely exponential $\exp(-i\Gamma_1t/\hbar)$ (dashed line) are plotted. Note also, since $R =0.91$, of the order of unity, that
the frequency of oscillations around the exponential-nonexponential transition is much more reduced than in the previous case.

\begin{figure}[!tbp]
\rotatebox{0}{\includegraphics[width=3.3in]{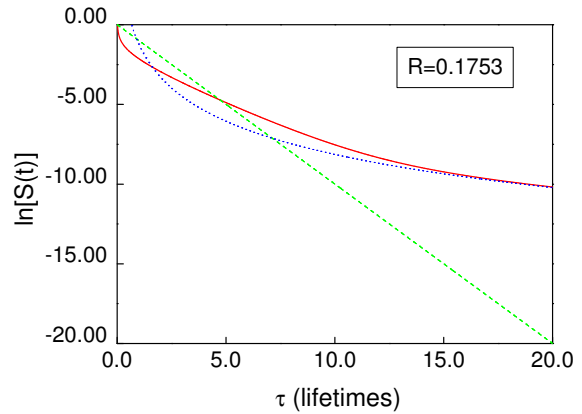}} \caption{`online color'.  Full nonexponential behavior of the survival probability
$S(t)$ (solid line), for a double-barrier system as a function of time. Here $\tau_1=1.64$ fs. The purely exponential behavior for $S(t)$
(dashed line), and the long time asymptotic behavior of $S(t)$ (dotted line), are included for comparison. See text.} \label{nonexp}
\end{figure}

\begin{figure}[!tbp]
\rotatebox{0}{\includegraphics[width=3.3in]{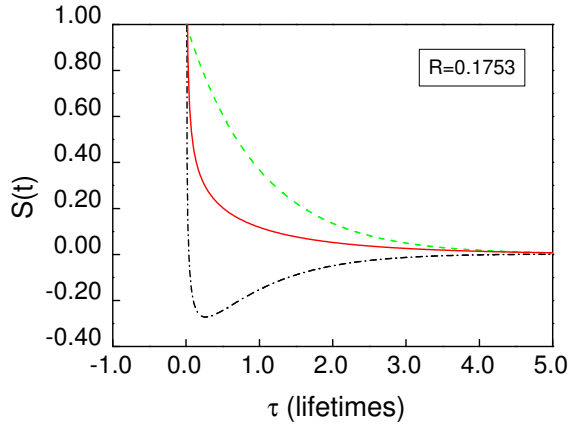}} \caption{`online color'. Survival probability $S(t)=S(t)_{exp}+S(t)_{non}+S(t)_{int}$
(solid line) for a few lifetimes to show that $S(t)_{int}$ (dashed-dot line) causes the deviation from the exponential contribution $S(t)_{exp}$
(dashed line) in the pre-asymptotic regime for the same case of Fig. \ref{nonexp}. See text.} \label{interf}
\end{figure}

Now, it turns out that by considering systems with still smaller values of $R$, leads to the regime where that decay proceeds entirely in a
nonexponential fashion.  We have found that this occurs for values $R \lesssim 0.3$ \cite{rvalue}. Figure \ref{nonexp} illustrates an example of
this regime. The potential parameters for the barrier widths and barrier heights remain the same as in the previous case, $b=1.0$ nm and
$V=0.23$ eV, but the well width takes now the value $w=1.5$ nm. Note that since each monolayer of semiconductor material has a thickness of
about $0.25$ nm \cite{barnham}, each barrier and the well involve, respectively, $4$ and $6$ monolayers. For this system the resonance
parameters are: $\varepsilon_1=0.07025$ eV and $\Gamma_1=0.40075$ eV. Also $R=0.1753$ and $\tau_1=1.64$ fs. Here the real part of the complex
pole, $\alpha_1=0.0491$ is still larger than the corresponding imaginary part $\beta=.03532$, which means that the pole is proper and provides
an exponentially decaying contribution \textit{i.e.}, Eq.\ (\ref{12d}). Although the resonance width is much broader than the resonance energy,
the double-barrier system is  still able to trap the particle. One sees that the survival probability $S(t)$ (solid line) exhibits a behavior
that departs from the purely exponential behavior $\exp(-i\Gamma_1t/\hbar)$ (dashed line) along the full time span. Note that the long time
regime, $S(t)_{non})$ (dot line), becomes the dominant contribution only after 15 lifetimes. Hence, one may ask what originates previously the
deviation of $S(t)$ from the exponential behavior. The answer follows by inspection of the expression for $S(t)$ given by Eq.\ (\ref{12c}): the
deviation from exponential behavior is due to the interference term $S(t)_{int}$.
This is illustrated in Fig. \ref{interf}. The interference term  $S(t)_{int}$ (dashed-dot line), given by Eq.\ (\ref{12f}), adds up a negative
contribution to the exponential decaying  contribution $S(t)_{exp}$ (dashed line) given by Eq.\ (\ref{12d}), to yield a nonexponential behavior
of the survival probability $S(t)$ (solid line) in a time span that for larger values of $R$ is usually dominated by the exponential term.
\begin{figure}[!tbp]
\rotatebox{0}{\includegraphics[width=3.3in]{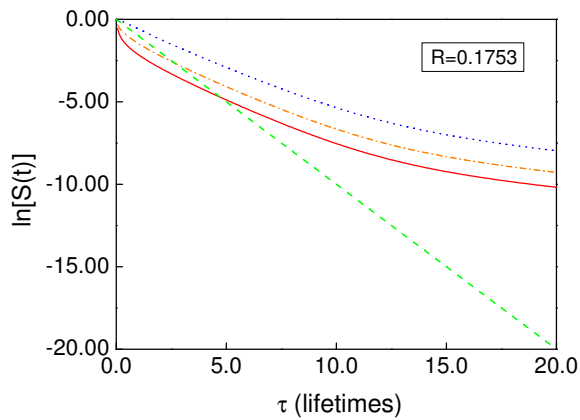}} \caption{`online color'. The survival probability $S(t)$ for different initial states:
infinite box (solid line), exact single resonance (dot-dashed line) and resonant function $u_1(x)$ (dotted line). For comparison, the purely
exponential behavior is also plotted (dashed line). See text.} \label{comparison}
\end{figure}

As pointed out before, in the above calculations we have considered as initial state a box solution of  well width $w$. We have examined
different choices of the initial state to see how the nonexponential behavior of the survival  probability is affected. We have found some
quantitative differences but the nonexponential behavior remains unaffected. The differences arise from the distinct values of the expansion
coefficients as illustrated in Fig. \ref{comparison} for $R=0.1753$, where we compare:\,(a) the box initial solution (solid line) $[{\rm Re}\,(
C_1{\bar C}_1 )=0.611]$, (b) the analytical exact single level resonance formula\cite{gcrr01} (dash-dotted line) $ [{\rm Re}\, (C_1{\bar C}_1)
=1.0]$, and, (c) the case where the initial state is the resonance function $u_1(x)$ along the internal region of the structure (dotted line) $
[{\rm Re}\,( C_1{\bar C}_1)=2.070]$.

We believe that a possible way to test our results for the full time nonexponential behavior of quantum decay, is by means of an experimental
setup analogous to that used by Sakaki \textit{et al.} \cite{sakaki} where a laser is used to create electron-heavy-hole pairs in the quantum
well of the double barrier. For thin barriers, as in the example discussed here, these authors showed that the decay process is dominated by
tunneling escape compared with the competing radiative recombination process. The decay rate of electrons is then measured indirectly by
analyzing the time-resolved photoluminescence. What is relevant here is that the value of $R =\varepsilon_1/\Gamma_1 \leq 0.3$. Clearly these
values of $R$ may be designed in other artificial quantum structures as in the decay of trapped atoms by lasers \cite{raizen}.

On completing this work it came to our notice a very recent work by Rothe \textit{et.al.} \cite{rothe}, where it is reported the long-awaited
experimental verification of the deviation of the exponential decay law at long times. This has been achieved by measuring luminescence decays
of dissolved organic materials. A distinctive feature of this work is that the small value of $R$ is induced by a local solvent environment. In
this respect this work differs from our approach which refers to the decay of an isolated system. It is to be expected that this experimental
work will stimulate further research in this area.

\section{Concluding remarks}

In summary, we have found  that the full time nonexponential behavior of the survival probability may be also characterized by three regimes:
(a)\, A first regime, encompassing a small fraction of the lifetime of the system, that is dominated by the short-time behavior and the high
resonance contributions to the survival probability;\,(b) A second regime, dominated by the interference  contribution between the exponential
and the nonexponential terms to the survival probability; \,(c) A third regime that is dominated by the long time asymptotic nonexponential
contribution to decay. In fact, (a) and (c) are regimes that are  present in general in any decaying system. The nonexponential behavior of
decay in stage (b) appears in systems  with a small value of the parameter $R$ in the range $ 0< R \lesssim\ 0.3$. For larger values of $R$ this
regime corresponds to the usual exponentially decaying behavior. Our approach possesses a general character for decay in quantum systems, and
therefore, it may be applied  to study the transition from exponential to nonexponential decay, and in particular the purely nonexponential
regime, in other suitable designed artificial quantum structures.

\begin{acknowledgments}
The authors thank Gonzalo Muga for useful discussions. They acknowledge partial financial support of DGAPA-UNAM under grant No. IN108003. J. V.
also acknowledges  support  from  10MA. Convocatoria Interna-UABC under grant No. 184 and Programa de Intercambio Acad\'emico 2006-1, UABC; and
G. G-C., from El Ministerio de Educaci\'on y Ciencia, Spain, under grant No. SAB2004-0010.
\end{acknowledgments}


\begin{thebibliography}{16}
\bibitem{threshold} Here and in the rest of the discussion we take the threshold energy $E_{th}=0$. Otherwise,
$R=(\varepsilon_r-E_{th})/\Gamma_r$.
\bibitem{rvalue} This is slightly smaller than the estimate given in Ref. \onlinecite{jittoh}.
\end{thebibliography}
\end{document}